\documentstyle{article}
\begin{document}
\large
~~~\\
\vskip 1.5 cm
\begin{flushleft}
      {\bf\Huge
          Fundamental Physical Constants\\
          and the Principle of Parametric Incompleteness} \\
\end{flushleft}
\vskip 0.5 cm

\begin{flushright}
\begin{tabular}{p{5.5in}}
{\bf\Large S.S. Stepanov}\\
Dnepropetrovsk State University\\
e-mail: steps@tiv.dp.ua\\
~\\
\hline
\normalsize
\it
The principle which allows to construct new physical theories 
on the basis of classical mechanics by reduction of the number of
its axiom without engaging new postulates is formulated. The arising
incompleteness of theory manifests itself in terms of theoretically
undefinable fundamental physical constants $\hbar$ and $c$.
As an example we built up a parametric generalization of relativistic theory,
where the Hubble Law and the dependence of light velocity on time
are obtained.
\\
\hline
\end{tabular}
\end{flushright}

\vskip 1 cm

                  \section{INTRODUCTION. QUESTIONS.}

It is hard to overestimate the importance of fundamental constants
$\hbar$ and $c$ in contemporary physics. They define the structure of
the theory's basic formulas so that these formulas can be transformed into
corresponding relations of classical physics by fixing limit value of the
constants $\hbar = 1/c = 0$. Their numerical value set scale of the
phenomena where respective corrections to the classical mechanics
become essential. They have become so common that sometimes are
"forgotten" when working in the unity system with $\hbar=c=1$.

Nevertheless there are many questions connected with fundamental
constants where full answers are not known yet. Some of them are
listed below:

\begin{itemize}
\item  Why physical theory is uncapable of calculating constants without
resorting to the experiment?
\item  Why fundamental constants emerge in more general physical theories
such as quantum and relativistic mechanics but are absent in classical
physics?
\item Are fundamental constants, which differ from  $\hbar$ and $c$, and
relating to them generalization of classical mechanics possible?
\item Is the set of possible fundamental constants finite?

\item  Do physical "constants" depend on time?
\end{itemize}

There is a great number of constants in physics thus first of all it is
necessary to point out what we mean by using the term "fundamental
constants".

For instance, speaking about the constant $c$ one usually uses the
term {\it "light velocity"}. At the same time two essentially different
constants are used under the same name: the velocity of propagation of
electromagnetic waves in vacuum $c_{em}$ and invariant velocity $c$ defining
the structure of the relativistic theory. The fact that the value of the
electromagnetic waves velocity $c_{em}$ is equal to the value of invariant
velocity $c$ is a property of one of the existing interactions while the
constant $c$ is build-in any form of substance. In particular, in order
to measure the value of invariant velocity there is no need to make
electrodynamic experiments. It would be enough to measure a speed of
any particle in two frames of reference and get the value from the formula
of speed addition: $u' = (u+ v) / (1+uv/c^2)$.
Even if photon had a different from zero mass and there were no other
massless particles the theory of relativity with the constant $c$
wouldn't have changed.

Consequently, the constant $c$ is a fundamental physical constant, and
$c_{em}$ which has the same numerical value is only a parameter of one of the
interactions (related to the value of the photon mass) and is not
fundamental.

In general, present day's physics consists of three closely
related and intersecting parts:

\vskip 0.5 cm

\begin{tabular}{lll}
MECHANICS    & (classical,quantum,relativity...) &  $\hbar$,$c$,..  \\
INTERACTIONS & (electroweak, strong,...)         &  $e$, $c_{em}$, ..\\
STRUCTURE    & (electron, muon, atom,...)        &  $m_e$, $m_p$,..\\
\end{tabular}

\vskip 0.7 cm

MECHANICS sets laws that are applied to any structural unities and relations
between them. It is the basis of two other parts of the
physics building. For instance, requirement of relativistic invariance
and unitarity restricts a class of possible interactions. One and the
same INTERACTION can be realized between different
STRUCTURAL unities, the variety of which defines diversity of
manifistations of our World.

Correspondingly, physical constants can also be divided into three classes
(similar classification is given in \cite{Levy}).
Let's hereafter assume to name
fundamental only those constants which define formulae
structure of theories applicable to all the forms of substance and kinds
of interactions, i. e. those constants which define properties of
MECHANICS. In that way now we know three fundamental
constants: Plank's constant $\hbar$, fundamental velocity  $c$ and,
apparently, gravitation constant $G$. Electron charge, masses of
elementary particles and other important parameters are not
fundamental in the sense stated above.

In classical mechanics fundamental constants are absent. More exactly,
their value is trivially fixed ($0$ or $\infty$).
Gravitation constant $G$ as well as
light velocity are present in classical physics but it obtains its
fundamental sense only in contemporary theories of space and time.


\vskip 1 cm
                  \section{AXIOMATIC BASIS OF THEORIES.}

By our opinion the resolution of the questions given in the introduction
is connected with the axiomatic analysis of the grounds of physical theories.
In mathematics the axiomatic method has been used since the Euclid times
but serious attention has been paid to the questions of axiomatic only after
the appearance of non-Euclidian geometry and paradoxes in the theory of
manifold. The metamathematics with the help of which an axiomatic
structure of different parts of mathematics is analised was created by
Gilbert and other mathematicians. Any axiomatic system of a theory
must possess the following features: independance, incontradictionity
and completeness \cite{Gilb}.

In particular, completeness means that any statement of a theory must
be proved or denied with the help of initial axioms. So classical physics
with respect to fundamental constants is complete while
the theory of relativity is not
because such statements as $c = 300,000$ $km/sec$ can neither be proved
or denied deductively (of course, without making a corresponding experiment).

For more then two centuries there were attempts to prove the fifth
axiom of Euclids parallel geometry. I. e. there was a suspicion that
Euclidian axiom system is not independant. Trying to prove the
axiom about parallels there were derived many theories which did not
depend on that axioma - "perfect geometry" according to Boyai's
terminology. A new theory has been created - non-Euclidian geometry. Unlike
Euclidian geometry this theory possesses a new constant $R$-radius of space
curvature, the value of which cannot be defined from initial axioms and
with $1/R \to 0$ these formulas transfer into respective theorems of
Euclidian geometry.

If in the theory there is a constant which value is impossible to derive from
initial axioms we shall call it {\it\bf parametric incompleteness of the
theory}. Why does it occur?  Obviously because a
new, shortened system of axiom contains less information then initial
one. A decrease of information results in some incompleteness of the
theory's conclusions. This incompleteness can be minimal, i. e. all the
functional relations of a theory can be obtained from initial axioms and
only final set of constants will remain undefineable.

Following Einstein, there is rather common but axiomatically incorrect way to
build up relativistic theory basing on two postulates: the principle of
relativity and the principle of constancy of light velocity.
It looks like that to build up the theory of relativity it is necessary to
add a new axiom of "$c_{em}$-invariance " to the classical Galilei theory
of relativity. It is rather wrong. Lorentz transformations and all
the theory of relativity formulas derived from axioms
of classical mechanics in which
absoluteness of time is denied. And the statement $c=inv$ is a theorem
of a theory but is not its initial statement.

The fact that Lorentz transformations can be obtained from simple
group considerations without using the second Einstein postulate was
known in 1910 yet owing to the works of Ignatowsky, Frank and Rothe
(\cite{Pauli},\cite{Berzi} for references).
In spite of the fact that this simple and beautiful result
has been reprinted in literature several times for the last ninety years it
has not become a possession of text-books though. Some latest works
of this line can be found in \cite{Berzi} -\cite{Nishikawa}, references
for earlier sources in \cite{Berzi}.

As it is known there is a definite analogy between geometry and
the theory of relativity.
Let's consider a velocity space, i. e. a three-dimensional space,where each
point represents one or another inertial system. The vector
between two points of this space corresponds to the vector of relative
velocity of two systems. The principle of relativity or inertial system
equality according to the language of geometry means that the
velocity's space is gomogeneous and isotropic, i. e. there are no
privileged points (frames of reference) and directions within this
space.

If we suppose that the velocity space has a Riemannan structure, we
would see that there are {\it only three} possibilities follow from the
principle of relativity: a plane space, a space of negative and positive
constant curvature. The first one corresponds to Galilean rule of
summing up velocities and to classical mechanics, the second one
corresponds to the theory of relativity, the third one, probably,
can't be realized in nature though it is formally used in Euclidian theory
of field.

The velocity space with constant negative curvature is a Loba\-chevsky
space. The curvature of this space corresponds to the value of
fundamental velocity $c$. In this way, the theory of relativity and classical mechanics are
already contained in the principle of relativity and Lorentz
transformations can be obtained without using a postulate $c=inv$. In
this regard as well as in Euclidian geometry there is a fundamental
constant $c$ associated with a decreased number of initial axioms (or
information they contain). A giving of the axiom of time absolutivity up
makes the theory of relativity a parametrically incomplete.


\vskip 1 cm

\section{PARAMETRICAL INCOMPLETENESS \\ AS A PRINCIPLE OF ACCORDANCE.}

In August, 1900 David Guilbert formulated his famous 23 problems on
the II International congress of mathematicians. The sixth problem was
a purpose of physics axiomatisation. Almost immediately after that two
new mechanics had been constructed: quantum and relativistic.
The crucial features of these theories are two fundamental constants
$\hbar$ and $c$.

Within any mechanics one might build up his own axiomatic systems
from which all the relations of the theory can be derived. But it is not a
satisfactory solution of Guilbert's sixth problem. The existance of
several systems of axioms breaks integral structure of physics. Besides,
this does not allow to obtain new physical theories by axiomatic means.

After axiomatic definitions of basic concepts (Space, Time, Mass,
State) classical mechanics becomes rather formal mathematical theory,
the axiomatic system of which must satisfy the conditions of
completeness, independence and incontradictivity.

Any decrease of information contained in axiomatic scheme of classical
mechanics leads to undefineable (unconcludeable)
parametres, functions, etc. There exist such minimal information
simplifications of mechanics axioms when only final number of
constants becomes undefineable and all the functional relations are derived
straightforwardly. Thus, on the basis of classical mechanics it is, probably,
possible to form a set of parametrically incomplete theories
deductively. In these theories a part of fundamental physical constants
will be played by constants which origin is related to a decrease
of the initial information containing in axioms of classical mechanics. At
the limit values of these constants we will obtain a classical theory
again.

{\it
In this way we have, so to say, a converse principle of correspondence.
One may obtain classical mechanics not
from the quantum theory or the theory of relativity
but obtain new fundamental theories from classical physics.
}

Using the language of algebra,
it is necessary to look for all the
possible stable deformations of algebraic structures of classical physics.
Parameters of deformations inevitably emerging in this connection
will be fundamental constants of the new physics. We already know
three of such deformations. The construction of other physical
theories possessing a corresponding parametric incompleteness looks
is more than interesting prospect.

It may happen that despite all the variety (probably infinite)
of objects and
their interactions in our World there exists a finite number of
fundamental physical theories (mechanics) which are parametrically
incomplete and can be derived from axioms of classical mechanics.

\vskip 1 cm
             \section{PROJECTIVE THEORY OF RELATIVITY.}


As an example of an application of Parametrical Incompleteness Principle
 let's consider some axioms of classical physics related to
the principle of relativity. We are interested in a manifest relations between
coordinates $x$ and time $t$ of two observers in different inertial frames
$S$ and $S'$:
\begin{equation} \label{Coord}
  x'=f(x,t),  \quad  t'=g(x,t)
\end{equation}
 In classical mechanics we requires the following to be realized first of all:

\begin{quotation}
 {\rm\bf Axiom I.~} \it
Transformations of coordinates and time are continuous, differentiable and
one-valued functions.
\end{quotation}
This requirement is very natural, and though it narrows the class of possible
functions of transformation it nevertheless leaves this class rather wide.
Each additional axiom reduces arbitrariness of choice of transformations (\ref{Coord})
unless this arbitrariness appears parametric or disappear completely
(in classical mechanics)
\begin{quotation}
 {\rm\bf Axiom II.~} \it
If some body moves uniformly in the system $S$ it's movement in
the system $S'$ will also be uniform.
\end{quotation}
The axiom (II) is actually a definition of inertial systems and time.
"The time is defined so that moving to be simple."
\cite{Misner}.

Despite rather common character of Axiom (I), (II), the functional dependence of
transformation can be fixed completely.
\begin{equation}\label{Coord2}
  x'=\frac{Ax+Bt+C}{ax+bt+c},  \quad  t'=\frac{Dx+Et+F}{ax+bt+c},
\end{equation}
leaving  nine parameters $A,B,C,D,E,F,a,b,c$ (see Appendix1)

We should notice that linear-fractional transformations with the same
denominator are well known as projective transformations in Geometry.
These are more common transformations when all the straight lines transfer
into straight lines \cite{Klein}.
That is an essence of the Axiom (II) in a two-dimensional space ($x,t$)
\begin{quotation}
 {\rm\bf Axiom III.~}\it
 At the moment of time $t=t'=O$ origins of systems coincide: $x=x'=0$
\end{quotation}

\begin{quotation}
 {\rm\bf Axiom IV.~}\it
 All the points within the frame of each observer are fixed.
\end{quotation}

\begin{quotation}
 {\rm\bf Axiom V.~} \it
 If the point of the system $S'$: $x'=0$ moves at a speed of $v$ relatively
 $S$ the point of the system $S$: $x=O$ moves at a speed of $v' = -v$ relatively
 $S'$
\end{quotation}

The Axiom (III) physically means the possibility of local, simultanuos experiment
for both observers which would allow to fix origin of coordinates in
space and time.

As far as the point $x'=O$ is stationery in the system according to (IV),
it can be derived from the axiom (II) that it moves uniformly and linearly
at some speed $v$ relatively $S$: $x=x_0+vt$. In this regard (III)
the following can  be derived: $x_0=0$

Analogically the beginning of the system $S$ relatively $S'$ moves at a speed of
$v':~~ x'=-v't'$. In this connection the axiom (V) is introduced.
Comparing to previous axioms the fifth axiom is not that "obvious".
Moreover, this axiom cannot be realized in absolute (ether) theories where
relativistic principle \cite{Sexl} -  \cite{Selleri2} is violated.
We should also point out that it can be considered as a theorem which derives
from linearity of transformations (\ref{Coord}), isotropy of space and principle
of relativity \cite{Berzi}.
Let's consider that (V) is an independent axiom allowing the observers in $S$ and
$S'$ to coordinate unit of measurement of time(or length) by mutual agreement
about the equality of relative speed measured by them.

Taking into consideration Axioms (I)-(V) we can write the following coordinate
transformations (\ref{Coord}) :
\begin{equation}\label{Transf1}
  x'=\frac{\gamma(v)(x-vt)}{1+a(v)x-b(v)t},  \quad
  t'=\frac{\gamma(v)(t-\sigma(v)x)}{1+a(v)x-b(v)t},
\end{equation}
where $\gamma(v),\sigma(v),a(v),b(v)$ - are arbitrary functions of relative speed.

The requirement of relative isotropy of space:
\begin{quotation}
 {\rm\bf Axiom VI.}\it
At inversion of axles of coordinates of both systems $S,S'$: $x\to -x$ and
$x'\to -x'$ of transformations (\ref{Coord}) are invariant.
\end{quotation}
is valid only in case if $\gamma(v), b(v)$ are even functions and
 $\sigma(v), a(v)$ -is odd.

\vskip 0.2 cm

The seventh axiom is the key in axiomatic of the relativistic theory
It expresses the principle of relativity and equality of inertial systems.

\begin{quotation}
 {\rm\bf Axiom VII.}\it
 There exist at least three equal inertial systems moving with arbitrary speeds.
\end{quotation}

If $X$  means the vector $(x,t)$, and $X'=\Lambda (X,v)$ the matrix notation
of transformation (\ref{Transf1}), for all frames $S_1,S_2$ and
$S_3$ we have:
$X_2=\Lambda (X_1,v_1)$, $X_3=\Lambda (X_2,v_2)=\Lambda (X_1,v_3)$,
so:
\begin{equation}
  \Lambda(\Lambda(X_1,v_1),v_2)=\Lambda(X_1,v_3), \quad  \forall X_1,v_1,v_2
\end{equation}
This functional equation gives the following relations between coefficients
of transformation (\ref{Transf1}):
\begin{equation}\label{Res1}
  \left\{ \begin{array}{l}
  \gamma_3 = \gamma_1\gamma_2(1+\sigma_1v_2) \\
  \gamma_3 = \gamma_1\gamma_2(1+v_1\sigma_2)
  \end{array}\right.
  \left\{ \begin{array}{l}
  v_3\gamma_3 = \gamma_1\gamma_2(v_1+v_2) \\
  \sigma_3\gamma_3 = \gamma_1\gamma_2(\sigma_1+\sigma_2)
  \end{array}\right.
  \left\{ \begin{array}{l}
  a_3 = a_1+a_2\gamma_1+b_2\gamma_1 \sigma_1 \\
  b_3 = b_1+b_2\gamma_1+a_2\gamma_1 v_1,
  \end{array}\right.
\end{equation}
where $\gamma_3=\gamma(v_3)$ etc. The first system of equations
in (\ref{Res1}) allows us to find a function $\sigma(v)$:
\begin{equation}\label{Const1}
 \frac{\sigma(v_1)}{v_1} = \frac{\sigma(v_2)}{v_2} = \alpha = const
\end{equation}
As far as speeds $v_1$ and $v_2$ are arbitrary independent values, $\alpha$
is some constant which is fundamental and the same for all the inertial
systems.
In our World it equals inverse square of "light velocity": $\alpha = 1/c^2$.
It is impossible to fix numerical value of this constant without additional
axioms.
So we have a manifestation of the principle of parametrical incompleteness and
appearance of parametrically incomplete generalization
of classical mechanics which
is known as special theory of relativity.

The requirement of equality of inertial systems together with
the axiom (V) leads to the inverse transformation:
$X'=\Lambda(X,v) \Rightarrow X=\Lambda(X',-v)$,
which results the following (taking into account axiom (VI)):
\begin{equation}\label{Res2}
  \left\{ \begin{array}{l}
  \gamma^2 =1/(1-v\sigma)  \\
  (\gamma-1)a = b\gamma\sigma\\
  a\gamma v  = (1+\gamma)b.
  \end{array}\right.
\end{equation}
The first equation of the system (\ref{Res2}) gives a manifest value of
Laurence factor $\gamma$. The equations of the systems (\ref{Res1}),(\ref{Res2})
lead to a functional equation (See Appendix 2):
\begin{equation}\label{Const2}
 \frac{a(v_1)}{v_1\gamma(v_1)}=\frac{a(v_2)}{v_2\gamma(v_2)}=\lambda = const,
\end{equation}
where $\lambda$ - is a new fundamental constant which is the same
for all the inertial systems.
Until now we used only one space dimension. Let's change the axiom (III) for:
\begin{quotation}
 {\rm\bf Axiom III'.~}\it
 At the moment of time $t=t'=0$ of the planes $(y,z)$,  $(y',z')$,
 which satisfy the equations
 $x=0$, $x'=0$,are parallel. Axles $y,y'$ and $z,z'$ are mutually parallel
 as well.
\end{quotation}
It is easy to obtain formulae for coordinates which are transverse
to moving.
So we finally have:
\begin{equation}\label{Transf2}
    t'=\frac{\gamma(t-\alpha vx)}{1+ax-bt}, \quad
    x'=\frac{\gamma(x-vt)}{1+ax-bt},        \quad
    y'=\frac{y}{1+ax-bt},                 	\quad
	z'=\frac{z}{1+ax-bt},
\end{equation}
where
\begin{equation}
    \gamma=\frac{1}{\sqrt{1-\alpha v^2}},  \quad
	a=\lambda v \gamma,  \quad  b=\frac{\lambda}{\alpha}(\gamma-1).
\end{equation}
These transformations fully correspond to seven axioms formulated above.
The only arbitrary quantities are two constants  $\alpha$ and $\lambda$.
As far as these constants are the same for all the inertial reference systems
we can call them fundamental.

Obviuosly classical mechanics also follows the axioms (I)-(VII) but
it eliminates parametrical incompleteness by introducing two additional
statements:
\begin{quotation}
 {\rm\bf Axiom VIII.}\it~
 If the speeds of two particles are equal in one frame of reference
 they will be equal in another one.
\end{quotation}

\begin{quotation}
 {\rm\bf Axiom IX.}\it~
 If two events are simultaneous in one frame of reference they
 will be simultaneous in another one.
\end{quotation}
The Axiom (VIII) leads to condition $\lambda=0$, and (IX) - to $\alpha=0$.
The Axioms (I)-(IX) are independent and fully define transformation (\ref{Coord}).
With their help Galilei equations are derived:
\begin{equation}\label{Gal}
  x'=x-vt,  \quad  t'=t.
\end{equation}
Classical mechanics which contains Axioms (I)-(IX), is parametrically complete
theory. There are no undefinable physical constants. We may say that the system
(I)-(IX) is informationally complete.
If we exclude axiom (IX) in the independent system of axioms (I)-(IX) the quantity
of information will be decreased and we shall inevitably obtain some incompleteness
of the theory. But this incompleteness is limited only by indefineable constant
"$c$", i.e. it leads to a parametrically incomplete theory.	In this respect the
axiom (IX) maintains minimum amount of information. The Axiom (VIII)
(which omition
would lead to generalization of the theory of relativity and new parametrically incomplete
mechanics with some new fundamental constant) possesses the same property.
We should note that the omition of the second proposal of the axiom (III')
would lead to another parametrically incomplete generalization of classical
mechanics.
If $\lambda$ and $\alpha$ are non-zero we obtain a theory which generalizes the
relativistic theory. It is convenient to call it as
{\bf Projective Theory of Relativity}.
Obviously, there are possible four limited situations which can be realizad with
different scales of observed phenomena.
So with $\lambda=0$ we obtain ordinary Lorentz transformations
and with $\alpha\to 0$ - projective generalization of Galilei transformations:
\begin{equation}\label{Gal2}
    t'=\frac{t}{1+\lambda x v- \lambda t v^2/2},     \quad
    x'=\frac{x-vt}{1+\lambda x v- \lambda t v^2/2}
\end{equation}
As far as $\lambda c$  has a dimension of inverse length
any corrections of Lorentz
transformations can be detected only long later after initial synchronisating
experiment (Axiom III) or at long distances from origin of coordinates.

It seems natural that the decrease of the number of axioms demands
a serious reconsideration
of our intuitive concept of space and time.
First time this happened when a special theory of relativity appeared. 
We may suppose that relativity of notions
can be extended at a further building of
parametrically incomplete theories.


\vskip 1 cm
        \section{TRANSFORMATIONS FOR SPEED. \\ LIGHT VELOCITY.}


Transformations for speed can be obtained from space-time transformations of
(\ref{Transf2}) by standard way.
Defining $u_x=dx/dt,~u'_x=dx'/dt',~.. $ and taking differentials of
(\ref{Transf2}), we find:
\begin{eqnarray}\label{Speed}
  u_x'&=&\frac{u_x-v-(x-u_xt) b/\gamma}{1-\alpha u_x v + \lambda v (x-u_x t)},\\
  u_y'&=&\frac{u_y/\gamma+\lambda v(xu_y-yu_x)+(y-u_y t)b/\gamma}{1-\alpha u_x v + \lambda v (x-u_x t)}.
\end{eqnarray}

1.If we consider some point fixed in the system $S'$ ($u'_x=u'_y=0$),
then relatively to observer in $S$ it moves at $t=0$ with a speed of
\begin{equation}\label{Speed2}
  \vec{u}=\vec{v}+\lambda c^2 \left(1-\sqrt{1-\frac{v^2}{c^2}}\right) \vec{r},
\end{equation}
where $\vec{v}=(v,0,0),~ \vec{r}=(x,y,z)$.
Despite the fact that relatively to observer in $S'$ all the points of his system
are fixed (i. e. they have the same zero speed) $S'$ points have different speed
from the point of view of the reference system  $S$. Moreover, they
move away
in radial directions relative to the point displaced at $\vec{v}\gamma/b$
from the origin of coordinates. The system $S'$ seems as though it
expands from the point of view of stationary observer. In this way, relative
notion is not only a simultaneousness
of events but also a relative motionlessness of objects from the point of view of
different observers.

 2.We should note that the dependance of formulae (\ref{Speed}) on time $t$
 doesn't mean non-uniformity of free moving. If particle in the system $S$
 moves uniformly and lineary $x=x_0 + u_x t$, a moving in the system $S'$ will be
 uniform and linear (the system of axioms maintains it).
 But the speeds $u'_x$, $u'_y$ depend not only on the speeds
 $u_x$, $u_y$, but also on the location of particles at some fixed moment of time.
 For example, we have for $u'_x$:
\begin{equation}
  u'_x=\frac{u_x-v-\lambda c^2 x_0 (1-\sqrt{1-\alpha v^2})}{1-\alpha u_x v +\lambda v x_0}
\end{equation}
This property as well as relativity of mutual motionlessness is connected with
the fact that a projective transformation does not retain parallelism of
straight lines.

3.It is easiy to see that if the signal spreads in the system $S'$ at a speed of
$c$, this speed will not be equal to $c$ in the system $S'$.Moreover, it is
dependant on coordinates $\vec{r},t$. But we can examine that the following
value
\begin{equation}\label{SpeedLight}
  \vec{C}(\vec{r},t)=\frac{\vec{c} + \lambda c^2 \vec{r}}{1+\lambda c^2 t}
\end{equation}
is an invariant velocity. So for a one-dimension case this value transfers as a
velocity (\ref{Speed}):
\begin{equation}\label{SpeedLight2}
  C(x',t')=\frac{C(x,t)-v-(x-C(x,t) t) b/\gamma}{1-\alpha v C(x,t) +\lambda v (x-C(x,t) t)},
\end{equation}
where the same function(\ref{SpeedLight}) stands in the left and right part of
the firmula.
If we take derivative $u'_x$ on $u_x$ and equate it with zero we find that limit
can be achieved at $v=c$. The value $u'_x$ with any $u_x$ ¨ $v=c$
equals $C(x',t')$. In this way the value defined in (\ref{SpeedLight}), can
be considered as generalization of light velocity in the theory of relativity.

We should point out that the speed of light is $C(r,t)$ but not $c$.
The constant $c$ is fundamental physical constant and equal $C(r,t)$
only if $r=0$, $t=0$.

For an observer $S$ in $x=0$, light velocity $C(0,t)$ reduces after some time
(with $\lambda >0$).We should notice that not so long ago there appeared
works in which the idea of dependance of light velocity on time was used for
solution of cosmological paradoxes \cite{Moffat}- \cite{Albrecht2}.

We point out again that dependance of light velocity on time and coordinates
does not mean non-uniform moving of light signal.
Any light signal emitted from some point of space at time $(t_0,\vec{r}_0)$,
moves with a constant speed $C(\vec{r_0},t_0)$ along trajectory $\vec{r}=\vec{r_0}+\vec{C}(\vec{r_0},t_0) (t-t_0)$.
In particular, light impulse which passed the origin of coordinates at the moment of
synchronising experiment (Axiom III) moves along the trajectory
$x=ct$ at a constant speed $C(ct,t)=c$. The dependence of light velocity on
time and condition of its constancy along uniform motion trajectory leads to
the functional equation:
\begin{equation}\label{SpeedLight3}
  C\left(x_0+C(x_0,t_0)(t-t_0), t\right) = C(x_0,t_0),
\end{equation}
which must be true at any $x_0,t_0,t$. The simplest solution of equation
is (\ref{SpeedLight}).

\vskip 1 cm

\section{TRANSFORMATIONS BETWEEN OBSERVERS \\ OF ONE INERTIAL FRAME}


In all the formulae of the projective theory of relativity the point
$x=0,~t=0$ is privileged.
Actually, linear-fractional transformations are not considered because of their
non-homogeneity and consequently non-homogeneity of experiments in space and time.
We should notice that priveleged point $x=0,~t=0$, is obviously
associated
with an initial synchronising experiment. It is also isolated in space-time
in the theory of relativity  but it does not result in its non-homogeneity.

Another complication is connected with seeming non-equivalence of observers
inside one of the reference system. Let's consider two rest observers
which are situated at points $x=0$ ¨ $x=R$. A light signal emitted by the first
observer at a speed of $C(0,0)=c$ is received by the second observer in
$t=R/c$ time. But he is unable to reflect it with the same speed because in
this case it would be back in $t=2R/c$  and would have a speed which would be
more than a light velocity at that moment $c>C(0,2R/c)$.

To solve those problems one must consider transformations between two motionless
observers inside one inertial system. In other words,
we are looking for generalization of
transformations of translation in classical mechanics
\begin{equation}\label{Transl}
                   X=x-R, \quad Y=y; \quad T=t,
\end{equation}
which would be in accordance with formulae of the projective theory of relativity.

Let's go back to axiomatic method again. As far as we describe transformations
between two rest observers $(\vec{X},T)$ and $(\vec{x},t)$ in one inertial
system, we require axioms (I) and (II) to be fulfilled. Besides, transformation
$X=X(x)$ must not be dependant on time (motionlessness of observers) and must
be defined by one common parameter-relative distance $R$ which must be chosen the same
on agreement about the unit of measurement of distance. In this way, taking into
consideration that $X(R)=0$, $X(0)=-R$
more common linear-fractional transformations have the following form:
\begin{equation}\label{Transl1}
   X=\frac{x-R}{1-\sigma(R)Rx},
   \quad
   Y=\frac{\delta(R)y}{1-\sigma(R)Rx},
   \quad
   T=\frac{\alpha(R)+\beta(R)x + \gamma(R)t}{1-\sigma(R)Rx},
\end{equation}
where $\alpha(R),\beta(R),\gamma(R),\delta(R),\sigma(R)$ -are arbitrary functions
of the relative distance $R$.
Requiring  any $x_1,R_1,R_2$ complied with the
law of composition of transformations:
$x_3=X(X(x_1,R_1),R_2)=X(x_1,R_3)$,  we obtain:
\begin{equation}\label{TranslXY}
   X=\frac{x-R}{1-\sigma R x},
   \quad
   Y=\frac{y\sqrt{1-\sigma R^2}}{1-\sigma R x}.
\end{equation}
\begin{equation}\label{Transl4}
   T=\frac{\sqrt{1-\sigma R^2}t+\mu\vec{R}\vec{x} -(1-\sqrt{1-\sigma R^2})\mu/\sigma}{1-\sigma \vec{R}\vec{x}}.
\end{equation}
where $\sigma$, $\mu$ are fundamental constants.

It is easy to see that (\ref{TranslXY}) formally coincides with the formula of
addition of velocities in the theory of relativity.
In three-dimension  case transformation has the following form:
\begin{equation}\label{Transl3}
 \vec{X}=\frac{\vec{x}\sqrt{1-\sigma R^2}-\vec{R}+(1-\sqrt{1-\sigma R^2})\vec{R}(\vec{R}\vec{x})/R^2}
              {1-\sigma\vec{R}\vec{x}}.
\end{equation}
With $\sigma>0$ this is a Lobachevsky space in
Beltrami coordinates of tangent space.

For accordance with the projective theory of relativity we require
the light velocity $C(x,t)$
to be invariant for all the observers inside the inertial system.
Thus we shall write transformations for speed $U=dX/dT,~u=dx/dt$
\begin{equation}\label{Speed4}
  U=\frac{u\sqrt{1-\sigma R^2}}{1+\mu R u - \sigma R (x-ut)}
\end{equation}
and suppose $U=C(X,T),~u=C(x,t)$. We obtain a conclusion that invariance
is observed if $\mu=\lambda$ ¨ $\sigma=(\lambda c)^2>0$.

In this way we have the following physical situation: homogeneous and isotropic
coordinate space inside inertial reference system is a Lobachevsky space of constant
negative curvature. At the same time physical vector of direction is a vector
tangent to space. Physical length $R$ is connected with geometric $s$ by
equation $R=\lambda c \tanh(s)$.
Physical time is defined in such a way that free particles move uniformly and
lineary for all the rest observers.
We should notice that from the formula (\ref{Transl4}) it can be automatically
derived  that synchronising procedure looks in the following way:
Two observers which are at a relative distance $R$ from each other, for the point
lying at an equal distance from them
$x=-X=(1-\sqrt{1-(\lambda c R)^2})/(\lambda c)^2 R)$, are fixing equal time $T=t$.
In case of moving frame of reference it is necessary to use formulae
of the projective theory of relativity.

From the geometrical point of view the obtained transformations are six parametric
$(\vec{v},\vec{R})$ group transformation leaving forminvariant metric of special kind
in a flat Minkowsky space-time (See Appendix 3).
\vskip 1 cm

            \section{HUBBLE'S LAW. EVOLUTION OF THE UNIVERSE.}

An interesting conclusion from the formulae of previous sections arises
if Doupler's effect is analized within one inertial system.

Let's consider remote {\bf motionless} source of light with coordinates
$\vec{R}$ emitting light in the direction to observer which is situated at the
beginning of coordinates $x=0$. The light pulse emitted at the moment of time
$t_1$ according to observer's clock reaches it at the moment $t_2$. As far as the
speed of this signal is constant $C(R,t_1)=C(0,t_2)$ and it moves in the direction
towards the observer ($\vec{c}=-c \vec{R}/R$), we have the following
relations between $R,t_1,t_2$:
\begin{equation}\label{Habbl1}
 (t_2-t_1)c = R + \lambda c^2 R t_2.
\end{equation}
Let's suppose light pulses are emitted with some period $\tau_0=\Delta T_1$
and are received
with a period $\tau=\Delta t_2$. As far as the source's time  $T$ and the observer's
time $t$ are related according to (\ref{Transl4}), then at constant
$\vec{x}=\vec{R}$, $\Delta T$ equals $\Delta t/\sqrt{1-(\lambda c R)^2}$.
Thus the period of emission is $\tau_0=\Delta t_1/\sqrt{1-(\lambda c R)^2}$,
and having entered cosmological parameter of redshift we finally obtain:
\begin{equation}\label{Habbl2}
  1+z=\frac{\tau}{\tau_0}=\sqrt{\frac{1+\lambda c R}{1-\lambda c R}}.
\end{equation}
Interpreting the redshift according to Doupler's formula we obtain
Hubble's law:
\begin{equation}\label{Habbl3}
 \vec{V}=\lambda c^2 \vec{R}
\end{equation}

As we saw above the privelegness of the point $x=0,~t=0$ is connected with the initial
synchronising experiment for concording the units of length and time
by different observers. Nevertheless, space is homogeneious and  isotropic and
therefore all the observers should be equivalent.

But homogeneity in time is not quite usual.
First of all it's usualness is associated with the dependence of local light
velocity (or maximum possible physical velocity) on time:
\begin{equation}\label{SpeedLight01}
  C(0,t)=\frac{c}{1+\lambda c^2 t}.
\end{equation}
In the past light velocity turned into infinity with $t_0=-1/(\lambda c^2)$.
The fact that the "beginning of Time" $t_0$ is away from time $t=0$,
by the value proportional to fundamental constants, is also associated with
the procedure of defining units of measurement.

You may find the explanation of it in such example. Suppose, that some 
observer at the moment of time $t=0$ had defined the unit of length (ruler),
the unit of time (sec) and measured light velocity and received the result
$c=300000~km/sec$.
In some period of time $T$ his distant descendants found the unit of length
but they do not know the unit of time.
But it is known from "ancient manuscripts" that light passes $300000~km$ per
second. Descendants define respectively their clocks and
light velocity measured by
them from this moment equals:
\begin{equation}\label{SpeedLight02}
  C(0,T)=\frac{c}{1+\lambda c^2 T}
\end{equation}
There is used the same {\it numerical value} of  fundamental
constant $c$ in the formulae (\ref{SpeedLight01}),(\ref{SpeedLight02}).
But time $t$ and time $T$, are evidently different. It is easy to find
valation between of times $T=T(t)$. As far as $dx=C(T)dT=C(t)dt$, and $T(\tau)=0$,
we obtain by integrating the formula:
\begin{equation}\label{SheetT}
  T=\frac{t-\tau}{1+\lambda c^2 \tau},
\end{equation}
generalizing the notion of replacement of time. For descenders as well as for
their ancestors "the beginning of Time" is distant for the same value $(\lambda c^2)^{-1}$,
measured in different units of  time.

Now let's discuss applicability of theory consrtucted herein to the Real World.
As far as Hubble's effect is naturally described within
the projective theory of relativity
it would be natural to associate Hubble's constant $H=65~km/sec/Mps=6.7~10^{-11}~year^{-1}$
with the constant $\lambda c^2$.
In this case the change of light velocity in time would be the following:
\begin{equation}\label{SpeedLightChange}
  \frac{\Delta C}{C}=\lambda c^2 t = - 6.7~10^{-11} \frac{t}{year}.
\end{equation}

In this way we have the following cosmological model. We live in stationary
space of constant negative curvature $(\lambda c)^{-1}=const$. The evolution 
of our World is connected not with the expansion of Universe but with the  
variability of speed of light $C(r,t)$ with time and distance.
This leads to the observed redshift for radiation of distant objects. 

There is also another and not so radical possibility
of interpretation of Hubble's
extension. If $\lambda c^2 \ll H$, our theory gives only small
corrections to Hubble's law, when traditional extension contributes more
within the limits of the theory of Big Bang.
In this case the dependence of light velocity on time $C(x,t)$
can be considerable
only within little times from the beginning of Big Bang.



\vskip 1 cm
                 \section{CONCLUSION. QUESTIONS.}

So taking proposed relativistic theory as an example, we shown that
there is a possibility to obtain new physical theories generalizing classical
mechanics by reducing its axiomatic base. If the system of axioms is independent
we find incompleteness of the theory which can be minimal in some cases and can
cause the emerging of fundamental physical constants and respective theories.
Thus all the generalizations of classical mechanics are parametrically incomplete
theories, and fundamental constants are manifestations of this incompleteness.

Quantum theory with Plank's constant is also a parametrically incomplete theory.
The elements of its axiomatic building can be found at Dirac. It is shown in
\cite{Dirac} that Plank's constant emerges from natural classical 
requirement if to exclude the axiom of physical values commutation.
We should notice that one of the axiomatic directions of quantum mechanics -
a quantum logic, also arises from the idea of decrease of number of axioms.
In this case we consider the system of axioms of Boolean logic, where
the axiom of distributive \cite{QLog1} is omitted.
The resulted nondistributive lattice
happens to be isomorphic for some quantum-mechanical systems.

Deductive, preexperimental path of building new theories looks very attractive.
In this regard the following questions arise:
\begin{itemize}
\item  How to formalize the concept of information contained in physical axiom?
\item  Is there an efficient procedure of search of axioms,
which being omitted cause
minimal informational lose-parametrical incompleteness?
\item  Are parametrical generalizations of quantum mechanics possible?
\item  Is the amount of information contained in the system of axioms of classical
mechanics, and, therefore, a number of its possible generalizations and fundamental
constants, limited?
\item  Is the Nature limited by parametrical incompleteness?
\end{itemize}
Of course, the list of these questions can be continued. Anyway,
the investigation of axiomatic of physics is not only
academically interesting for it extenses our understanding of Nature,
but also it can lead us to constructive results
allowing experimental testing.


\begin{flushright}
\bf{APPENDIX 1.}
\end{flushright}

Let's consider arbitrary, independant, differentiating transformations of
the coordinate $x$ and time $t$:
\begin{equation}\label{Primxtx't'}
  x'=f(x,t),  \quad  t'=g(x,t).
\end{equation}
We require the system of coordinates $(x,t)$ and $(x't')$ to satisfy the
definition of inertial reference systems:
\begin{equation}\label{PrimIner}
      \frac{du}{dt}=0 \quad \Longrightarrow \quad \frac{du'}{dt'}=0,
\end{equation}
i.e. if a moving of a body is uniform in one system, it would be uniform in
another one.

According to definition the speeds in each system are $u=dx/dt$
and $u'=dx'/dt'$, thus:
\begin{equation}\label{PrimSpeed}
				u'=\frac{f_x u + f_t}{g_x u + g_t},
\end{equation}
where $f_x=\partial f(x,t)/\partial x$, etc.
Differentiating (\ref{PrimSpeed}) on  $dt'=(g_x u + g_t)dt$
and equating coefficients
of speed $u$ (taking into consideration its arbitrariness)
to be zero, we obtain
the system of differential equations:
 \begin{eqnarray}
  f_{xx} g_x &=& g_{xx} f_x   \label{PrimSys1}\\
  f_{tt}\,\, g_t &=& g_{tt}\, f_t   \label{PrimSys2}\\
  f_{xx} g_t + 2 f_{xt} g_x &=& g_{xx} f_t + 2 g_{xt} f_x  \label{PrimSys3} \\
  f_{tt} g_x + 2 f_{xt} g_t &=& g_{tt} f_x + 2 g_{xt} f_t  \label{PrimSys4}.
\end{eqnarray}

Let's introduce Jacobian of transformations different from zero (\ref{Primxtx't'})
$D=f_xg_t-f_tg_x$.
Taking its derivatives by on $x$ and $t$ with the help of (\ref{PrimSys1}) -
(\ref{PrimSys4})  we obtain equations:
\begin{equation}
     2 \frac{D_x}{D} = 3 \frac{f_{xx}}{f_x}= 3 \frac{g_{xx}}{g_x};  \qquad
     2 \frac{D_t}{D} = 3 \frac{f_{tt}}{f_t}= 3 \frac{g_{tt}}{g_t},
\end{equation}
which are easily integrated and give the following equations:
\begin{equation}\label{PrimSolv1}
      \frac{f_t}{f_x} = \frac{A(x)}{B(t)};  \qquad
      \frac{g_t}{g_x} = \frac{\bar{A}(x)}{\bar{B}(t)};  \qquad
      \frac{g_t}{f_t} = \frac{\bar{A}(x)}{A(x)};  \qquad
      \frac{g_x}{f_x} = \frac{\bar{B}(t)}{B(t)},
\end{equation}
where $A(x),B(t),\bar{A}(x),\bar{B}(t)$ - are arbitrary functions.
We should notice
that two last equations in (\ref{PrimSolv1}) are direct results of
(\ref{PrimSys1}),(\ref{PrimSys2}).

Let's multiple the equation (\ref{PrimSys3}) by $f_t$, and (\ref{PrimSys4}) by $-f_x$ and add.
Then with the help of (\ref{PrimSys1}), (\ref{PrimSys2}) we obtain differential equation
only for the function $f(x,t)$:
\begin{equation} \label{PrimDifF}
	 f_{xx} f^2_t + f_{tt} f^2_x = 2 f_{xt} f_x f_t.
\end{equation}

Taking $f_{xx}$ and $f_{tt}$ from the first equation (\ref{PrimSolv1}) and putting
it in (\ref{PrimDifF}), we obtain:
\begin{equation}
	 A'(x) = - \dot{B}(t) = \alpha.
\end{equation}
As far as $t$ and $x$ are independent arguments, $\alpha$ is an arbitrary
constant. The equations(\ref{PrimSys1})-(\ref{PrimSys4}) are symmetric
under replacement of $f$ for $g$, thus we have similar equations for
$\bar{A}(x)$ and $\bar{B}(t)$. Thus:
\begin{eqnarray}
       A(x) &=& \alpha x + \beta; \qquad   B(t) =-(\alpha t + \gamma); \nonumber \\
 \bar{A}(x) &=& \bar{\alpha} x + \bar{\beta}; \qquad  \bar{B}(t) = -(\bar{\alpha} t + \bar{\gamma}),
\end{eqnarray}
where $\alpha,\beta,\gamma,\bar{\alpha},\bar{\beta},\bar{\gamma}$
- are constants which do not depend on $x$ and $t$.

Integrating the third and fourth equations (\ref{PrimSolv1}), we have:
\begin{equation}
  g(x,t)=\frac{\bar{A}(x)}{A(x)}f(x,t) + M(x) = \frac{\bar{B}(t)}{B(t)}f(x,t) + N(t)
\end{equation}
or
\begin{equation} \label{PrimSolv2}
  f(x,t)=(M(x)-N(t))/\left(\frac{\bar{B}(t)}{B(t)} - \frac{\bar{A}(x)}{A(x)} \right),
\end{equation}
where $M(x),N(t)$ - are arbitrary functions.
Putting (\ref{PrimSolv2}) in the first equation (\ref{PrimSolv1}), we have:
\begin{equation} \label{PrimSolv3}
 (\alpha x + \beta)M'(x) + \alpha M(x) = (\alpha t + \gamma)\dot{N}(t) + \alpha N(t) =  \sigma,
\end{equation}
where $\sigma$ - is arbitrary constant. The equations (\ref{PrimSolv3})  are
easily integrated and give for $M(x)$ and $N(t)$:
\begin{equation} \label{PrimSolv4}
 M(x) = \frac{\sigma x + \lambda}{\alpha x + \beta}; \qquad
 N(t) = \frac{\sigma t + \mu}{\alpha x + \gamma},
\end{equation}
which along with (\ref{PrimSolv1}) finally lead us to lineary-fractional transformations
with the same denominator.

\vskip 1cm
\begin{flushright}
\bf APPENDIX 2.
\end{flushright}

From the second equation of the system (\ref{Res2}) and first two systems (\ref{Res1})
we have:
\begin{equation}
 \frac{b_3}{a_3}=\frac{\gamma_3-1}{\alpha\gamma_3v_3}
 =\frac{\gamma_1\gamma_2(1+\alpha v_1v_2)-1}{\alpha\gamma_1\gamma_2(1+\alpha v_1v_2)}
\end{equation}
Let's divide the equations of the third system in (\ref{Res1}) by each other and
introduce designation: $F=a/(\alpha v\gamma)$. Then:
\begin{equation}
 \frac{ F_1(\gamma_1-1)\gamma_2 + F_2\gamma_1\gamma_2(\gamma_2-1+\alpha v_1v_2\gamma_2)}
      {F_1v_1+F_2(\gamma_2(v_1+v_2)-v_1)}
 =
 \frac{\gamma_1\gamma_2(1+\alpha v_1 v_2)-1}{v_1+v_2},
\end{equation}
where
\begin{equation}
 F_1(v_1\gamma_1+v_2\gamma_2-\gamma_1\gamma_2(v_1+v_2))
 =
 F_2(v_1\gamma_1+v_2\gamma_2-\gamma_1\gamma_2(v_1+v_2))
\end{equation}
or $F_1=F_2$

\vskip 1cm
\begin{flushright}
\bf APPENDIX 3.
\end{flushright}

It is easy to make sure that formulae of the projective theory of relativity
have the following
invariants:
\begin{equation}\label{PrilInvar}
   \frac{c^2 t^2 - x^2}{(1+\lambda c^2 t)^2}=inv,
   \quad
   \frac{c^2 t_1 t_2 - x_1 x_2}{(1+\lambda c^2 t_1)(1+\lambda c^2 t_2)}=inv.
\end{equation}
Using  (\ref{PrilInvar}) it  is easy to obtain forminvariant metric at
projective transformations:
\begin{equation}\label{Metrika1}
   ds^2
   =\frac{1-(\lambda c \vec{x})^2}{(1+\lambda c^2 t)^4}c^2 dt^2
   +\frac{2\lambda c^2 \vec{x} d\vec{x}dt}{(1+\lambda c^2 t)^3}
   -\frac{d\vec{x}^2}{(1+\lambda c^2 t)^2}.
\end{equation}
We can also make sure that metric (\ref{Metrika1}) remains forminvariant respective
transformations inside one reference system. In this way we have six-parametrical
group of transformations
($\vec{v},\vec{R}$).

Metric (\ref{Metrika1})can be written in the following way:
\begin{equation}\label{Metrika2}
   ds^2=-\frac{1}{(1+\lambda c^2 t)^2}
   \left(d\vec{x}-
   \frac{\vec{c}+\lambda c^2 \vec{x}}{1+\lambda c^2 t} dt
   \right)
   \left(d\vec{x}-
   \frac{-\vec{c}+\lambda c^2 \vec{x}}{1+\lambda c^2 t} dt
   \right).
\end{equation}
As far as $ds^2=0$ for spreading of light, we obtain the equation for light velocity
again $d\vec{x}/dt=\vec{C}(\vec{x},t)$.

By introducing new time
\begin{equation}
 \tau=\tau_0-\frac{1}{\lambda c^2}\frac{\sqrt{1-(\lambda c \vec{x})^2}}{1+\lambda c^2 t},
\end{equation}
we obtain Robertson-Walker's metric:
\begin{equation}\label{Metrika3}
   ds^2=c^2 d\tau^2 -a(\tau)^2(d\chi^2+\sinh^2 \chi~~d\Omega^2),
\end{equation}
where $a(\tau)=c|\tau-\tau_0|$ and spherical system of coordinates with radius
$\lambda c |\vec{x}|=\tanh \chi$ is introduced.
With the help of well-known transformations
\cite{Misner} we may transfer metric (\ref{Metrika3}) into Minkowsky metric.

\end{document}